\documentclass[11pt]{article}

\usepackage{jcappub} 
\usepackage[T1]{fontenc} % if needed
\usepackage{amsbsy}
\usepackage{color}
\usepackage{subcaption}

\usepackage{soul}

\usepackage{graphicx}
\usepackage[normalem]{ulem}
\usepackage{xurl}
\usepackage{enumitem}

\title{\boldmath Simulating the Universe from the cosmological horizon to halo scales}

\def\aap{\ref@jnl{A\&A}}
\def\jcap{\ref@jnl{J. Cosmology Astropart. Phys.}}

\author[a]{Thomas Montandon,}
\affiliation[a]{Laboratoire Univers \& Particules de Montpellier (LUPM),
CNRS \& Universit\'e de Montpellier (UMR-5299),
Place Eug\'ene Bataillon, F-34095 Montpellier Cedex 05, France}

\author[b,c]{Oliver Hahn,}
\affiliation[b]{Department of Astrophysics, University of Vienna, Türkenschanzstraße 17, 1180 Vienna, Austria}
\affiliation[c]{Department of Mathematics, University of Vienna, Oskar-Morgenstern-Platz 1, 1090 Vienna, Austria}

\author[d]{Cl\'ement Stahl}

\affiliation[d]{Université de Strasbourg, CNRS, Observatoire astronomique de Strasbourg, UMR 7550,
67000 Strasbourg, France}

\emailAdd{thomas.montandon@umontpellier.fr}
\emailAdd{oliver.hahn@univie.ac.at}
\emailAdd{clement.stahl@astro.unistra.fr}

\abstract{Ultra-large scales close to the cosmological horizon will be probed by the upcoming observational campaigns. They hold the promise to constrain single-field inflation as well as general relativity, but in order to include them in the forthcoming analyses, their modelling has to be robust. In particular, general relativistic effects may be mistaken for primordial signals, and no consensus has emerged either from analytical modelling nor from the numerical route, obstructed by the large dynamical range to be simulated. In this work, we present a numerical technique to overcome the latter limitation: we compute the general relativistic displacement field with the N-body relativistic code \texttt{gevolution} and combine it with the accurate Newtonian simulation \texttt{Gadget-4}. This combination leads to an effective simulation reproducing the desired behaviour at the level of the matter power spectrum and bispectrum. We then measure, for the first time in a simulation, the relativistic scale-dependent bias in Poisson gauge; at redshift $z=0$, we find $b_1^{\mathrm{GR}}=-8.1 \pm 2.8$. Our results at the field level are only valid in the Poisson gauge and need to be complemented with a relativistic ray tracing algorithm to compute the number count observable.}

\begin{document}

\maketitle

\section{Introduction}\label{sec:Introduction}
The most fundamental theory of gravity known and well tested today is General Relativity (GR). This theory is one of the main ingredient of the standard model of cosmology: the $\Lambda$CDM ($\Lambda$ Cold Dark Matter) model. Nowadays, one of the major issue in cosmology is the modelling of the Large-Scale Structure (LSS). Indeed, many experiments, currently operating or planned, such as DESI \cite{DESI:2013agm}, Euclid \cite{EUCLID:2011zbd}, the Vera Rubin Observatory \cite{LSSTScience:2009jmu} or SPHEREx \cite{SPHEREx:2014bgr}, will provide captivating fresh panoramas of LSS. These data will contain precious information on inflation, the evolution and content of the Universe and the theory of gravity. Extracting the maximum of information from these data requires however involved tools as LSS is fundamentally a nonlinear process which -- at least currently -- cannot entirely be described by analytical perturbative approaches. On large scales, in the CDM paradigm, all non-gravitational interactions can be neglected, and matter can be assumed to be perfectly cold. Moreover, the scales probed by the galaxy surveys have until now been small compared to the cosmological horizon. For all these reasons, there was no need to include horizon-scale relativistic effects in simulations. Hence, the state-of-the-art numerical simulations use the Newtonian gravity eg.~\texttt{Gadget-4} \cite{Springel:2020plp}, \texttt{PKDGRAV} \cite{Potter:2016ttn}, and \texttt{Ramses} \cite{Teyssier:2001cp}, see Ref.~\cite{Angulo:2021kes} for a review. 
 
As the volume probed by galaxy surveys increases, one approaches the cosmological horizon, and will even overpass it with future intensity mapping surveys at much higher redshifts \cite{Mellema:2012ht}. 
{Hence, a relativistic framework is needed to describe the dynamics at these scales. For example, we would need such a framework to describe photons and neutrinos in our Universe eg.~\cite{Euclid:2022qde}, as well as the propagation of light in a perturbed spacetime eg.~\cite{Euclid:2021rez,Rasera:2021mvk,Lepori:2022hke}. Another gain is that various extension of $\Lambda$CDM and extensions of GR are typically laid down in a covariant relativistic context, and their impact on LSS is then only probed in its linear (Newtonian) limit. While GR has 6 degrees of freedom, this restriction to the Newtonian scalar potential could ignore important constraining power hidden in the extra 5 coupled degree of freedom see Refs.~\cite{Christiansen:2024uyr,Christiansen:2024vqv} for a recent example.}

{
But it is important to stress that in the $\Lambda$CDM model, the description of the dynamics at these scales has not yet converged to a consensus despite the large literature on this topic, eg.~\cite{Matarrese:1997ay,Boubekeur:2008kn,Fitzpatrick:2009ci,Bonvin:2011bg,Jeong:2011as,Pajer:2013ana,Kopp:2013tqa,Yoo:2014sfa,DiDio:2015bua, DiDio:2014lka,Gallagher:2018bdl,Castiblanco:2018qsd,Erschfeld:2020blf,Maartens:2020jzf,Castorina:2021xzs,Elkhashab:2021lsk}. In particular, the complete theoretical modelling of the intrinsic relativistic bispectrum, i.e. the three point correlation in Fourier space, including radiation, is still missing. This is due to the highly complex and long equations that were derived and that should include many different effects such as the dynamical relativistic effects which is the subject of this paper, but also the so-called projection effects like redshift space distortions, Doppler and gravitational effects and lensing.}

{A good model is however crucial as it can be used to test the theory of GR and the $\Lambda$CDM model, and to constrain primordial non-Gaussianity (PNG) since relativistic effects and PNG can be degenerate \cite{Achucarro:2022qrl}. The amount of the so-called ``local'' PNG, which is one of the most considered in the literature, is often described by the parameter $f_{\rm NL}$ quantifying the quadratic correction to a Gaussian potential. Measurements of PNG at the level of $f_\text{NL}\sim 1$ are potentially within reach with the forthcoming LSS observational campaigns triggering many activities on the analysis of observational data \cite{eBOSS:2021jbt,Cabass:2022wjy,DAmico:2022gki,DESI:2023duv,Ivanov:2024hgq,Brown:2024dmv} and on designing large simulations to prepare and validate the data analysis pipeline \cite{Coulton:2022qbc,Coulton:2022rir,Anbajagane:2023wif,Adame:2023nsx,Jung:2024esv,Hadzhiyska:2024kmt}. 
However, these constraints on $f_{\rm NL}$ will rely on an accurate modelling of the large-scale galaxy power spectrum, where it is affected by PNG via the scale dependent bias, and the so-called squeezed limit of the bispectrum, limit where one mode is much smaller than the two others. The squeezed limit is particularly hard to model since it involves both large scales, where relativistic effects are important, and small scales where nonlinearities and radiation are important.}

{In the past years, we have proposed an alternative numerical route to try to settle this question: we resort to N-body codes taking into account radiation and relativistic effects in the initial conditions up to second order and in the dynamics \cite{Adamek:2021rot,Montandon:2022ulz}. These analysis have led to the conclusion that at redshift $2$, GR effects can reach an amplitude $1-10\%$ in the matter power spectrum and bispectrum. However, one fundamental ingredient is still missing, the second-order bias model. Indeed, we have used the state-of-the-art relativistic cosmological code}
\texttt{gevolution} \cite{Adamek:2016zes} {which} solves the 6 Einstein equations on a \textit{fixed grid} in the weak field regime, while being fully nonlinear for the stress energy tensor. While \texttt{gevolution} is extremely efficient, its main limitation is its lack of (adaptive) dynamic range making it impossible to resolve at the same time both the inner structure of halos along with the relevant large scales. To go beyond this limitation, many works have focused on including some relativistic effects in existing Newtonian codes. Most of these works have used the so-called N-body gauge developed in Ref.~\cite{Fidler:2015npa}, in which one can consistently describe the leading order general relativistic effects coupled to a Newtonian-like non-relativistic matter field. This way, linear radiation was included in the dynamics of \texttt{Gadget-3} in the code called \texttt{COSIRA} \cite{Brandbyge:2016raj}, and massive neutrinos were then included in \texttt{CONCEPT} \cite{Tram:2018znz} and \texttt{gevolution}, see~\cite{Euclid:2022qde} for a review. In Ref.~\cite{Adamek:2017grt}, the N-body gauge was extended beyond the scalar sector and implemented in \texttt{gevolution}. These linear methods were used to perform the \texttt{Euclid} flagship simulations. {Focusing on the scalar sector, Ref.~\cite{Fidler:2018geb} has developed a method allowing the interpretation of bias measured in Newtonian simulations in a GR context.} Finally, the merging of \texttt{gevolution} and its Newtonian counterpart \texttt{Gadget-4}~\cite{Springel:2020plp} was undertaken in \texttt{GrGadget}~\cite{Quintana-Miranda:2023eyn} while the extension of \texttt{RAMSES} to general relativity was performed in \texttt{GRAMSES}~\cite{Barrera-Hinojosa:2019mzo}.

In this article, we propose to combine the simulation outputs of \texttt{gevolution} and \texttt{Gadget-4} to go beyond the limiting grid size of \texttt{gevolution} and to perform for the first time simulations that can resolve both halos and scales close to the cosmological horizon. This combination is possible thanks to the developments of Refs.~\cite{Fidler:2015npa, Adamek:2017grt}, which allow \texttt{gevolution} to perform a Newtonian simulation. The advantage of our method is its simplicity, as it does not require any modification of the codes and can be applied {\it{a posteriori}} as post-processing. It is also nonlinear and goes beyond the scalar sector. In this paper, we will focus on the general relativistic dynamics, but the same method can be extended to eg. radiation and massive neutrinos. In section~\ref{sec:method}, we introduce briefly the results of Refs.~\cite{Fidler:2015npa, Adamek:2017grt} and our method to combine snapshots of \texttt{gevolution} and \texttt{Gadget-4}. In section~\ref{sec:setup}, we introduce the simulation setup used. In section~\ref{sec:stats}, we compare the summary statistics of the matter field in the different simulations and in section~\ref{sec:halos}, we study the properties of halos and measure for the first time in a simulation the scale dependence of the bias due to general relativistic effects. We also show some measurements of the halo bispectrum. In section~\ref{sec:conclusion}, we conclude.

\section{Method} \label{sec:method}
In this article, we propose a simple method to correct the small-scale gravity of \texttt{gevolution} with that from a purely Newtonian simulation. Specifically, we perform two simulations with the same initial random perturbation phases: one fully relativistic and one in the Newtonian mode of \texttt{gevolution} \cite{Fidler:2015npa, Adamek:2017grt, Adamek:2021rot}. The relativistic simulation is performed in the Poisson gauge. The Newtonian simulation however is performed in the so-called N-body gauge \cite{Fidler:2015npa}. This gauge was found to be the correct coordinate system to interpret Newtonian simulations in a relativistic context and at the linear level. Indeed, in the limit of late time, where radiation is negligible, the time perturbation of this gauge is vanishing. Moreover, the volume perturbation is set to zero, which means that the density perturbation of non-relativistic particles is equal to the density computed in a Newtonian N-body simulation. Finally, it can be shown that the linear equations of motion of non-relativistic particles in the absence of radiation, as well as the Poisson equation, are identical. For these reasons, the Newtonian \texttt{gevolution} simulation can directly be compared to the \texttt{Gadget-4} simulation as they are in the same coordinate system and share the same density perturbation and potential. 

If now we subtract the particle positions of the \texttt{gevolution} simulations
\begin{equation}\label{eq:xiGR}
    \boldsymbol{\xi}^{\mathrm {GR}} = \boldsymbol{x}^{\mathrm{GR}}_{\texttt{gev}} - \boldsymbol{x}^{\textrm{Newton}}_{\texttt{gev}}\,,
\end{equation}
we obtain a displacement field that contains two types of information. First, $\boldsymbol{x}^{\mathrm{GR}}_{\texttt{gev}}$ and $\boldsymbol{x}^{\textrm{Newton}}_{\texttt{gev}}$ are not in the same coordinate system. Hence, $\boldsymbol{\xi}^{\mathrm {GR}}$ can be interpreted as the spatial generator associated to the gauge transformation: Poisson$\rightarrow$N-body. Second, the relativistic simulation performed in the Poisson gauge also contains all the general relativistic effects, i.e. the cosmological horizon effects as well as the nonlinear coupling between the 6 degree of freedom. Remember that the Newtonian simulation has only one degree of freedom.
This subtraction cancels the ``Newtonian contribution'' which is contained in both positions $\boldsymbol{x}^{\mathrm{GR}}_{\texttt{gev}}$ and $\boldsymbol{x}^{\textrm{Newton}}_{\texttt{gev}}$ and it remains in $\boldsymbol{\xi}^{\mathrm {GR}}$ the pure relativistic contribution to the position of the particles. Note that, since \texttt{gevolution} is a particle mesh code, the field $\boldsymbol{\xi}^{\mathrm {GR}}$ is limited by the Nyquist frequency of the (uniform) simulation grid. 

Then, we perform a third simulation with \texttt{Gadget-4} again with the same initial random phases. We can now add $\boldsymbol{\xi}^{\mathrm {GR}}$ to the \texttt{Gadget-4} particle positions, such that 
\begin{equation}\label{eq:combined}
    \boldsymbol{x} = \boldsymbol{x}^{\mathrm{Newton}}_{\texttt{Gad}} +  \boldsymbol{\xi}^{\mathrm {GR}}\,.
\end{equation}
This combination allows us to perform a gauge transformation to the Poisson gauge, but also preserves the purely relativistic contribution and the small-scale Newtonian perturbation which are well resolved in the \texttt{Gadget-4} simulation thanks to the tree algorithm.

Note that we apply this transformation in a single post-processing step to a simulation (cf. also Ref.~\cite{Hahn:2016roq}). In principle, one could also envision a scheme, where the Newtonian small-scale correction is applied in every time step of a \texttt{gevolution} simulation. Since the corrections are mostly affecting large scales, it is however justified to neglect the back-reaction from the small-scale correction onto the large-scale evolution. {Note however that this is not the case for all general cosmological models. Here we are focusing on the stardard $\Lambda$CDM model to show that the method works in the simplest case. Indeed} 
As we will see in the following, working with integrated quantities already shows accurate enough results for the power spectrum and bispectrum, even correcting numerical uncertainties due to the discrete gradient computation of \texttt{gevolution}. Note also that Eqs.~\eqref{eq:xiGR} and \eqref{eq:combined} are in configuration space. Working in configuration space allows us to keep all the mode correlations which would not have been possible if we had combined the fields in Fourier space. Finally, note that \texttt{gevolution} is much more efficient (but less accurate), which means that the computational cost is largely dominated by the run in \texttt{Gadget-4}.

To make a fair comparison of the simulations combination, we propose in addition two alternative ways to compute the general relativistic displacement field $\boldsymbol{\xi}_{\mathrm {GR}}$. We use Lagrangian Perturbation Theory (LPT) following Ref.~\cite{Adamek:2021rot}, where it was developed iteratively (nLPT) in the discrete general relativistic case. For this paper, we will focus on the first ($n=1$) and second ($n=2$) order. We run the initial conditions generator \cite{Montandon:2022ulz} directly at the redshift of a given \texttt{Gadget-4} snapshot to obtain the particle positions using only $1$LPT or $2$LPT for the Newtonian ($\boldsymbol{x}_{\mathrm {nLPT}}^{\mathrm{Newton}}$) and the relativistic simulations ($\boldsymbol{x}_{\mathrm{nLPT}}^{\mathrm{GR}}$). The ``Lagrangian general relativistic displacement field'' $\boldsymbol\xi_{_{\mathrm{nLPT}}}^{\mathrm{GR}}$ can then be obtained by subtracting
\begin{equation}
    \boldsymbol\xi_{_{\mathrm{nLPT}}}^{\mathrm{GR}} = \boldsymbol{x}_{\mathrm{nLPT}}^{\mathrm{GR}} - \boldsymbol{x}_{\mathrm{nLPT}}^{\mathrm{Newton}}\,.
\end{equation}
Adding this displacement field to the particle positions of the corresponding \texttt{Gadget-4} snapshot should correspond to a $n^{\mathrm{th}}$-order gauge transformation. However, radiation and the discreetness of the method employed makes the interpretation of the errors subtle, as we will see in the following. The combination of \texttt{Gadget-4} with $1$LPT ($2$LPT) is simply labelled ``Combined $1$LPT'' (``Combined $2$LPT'').

\section{Simulation setup} \label{sec:setup}
We use a fiducial cosmology where the matter energy densities today are $\Omega_c = 0.263771$ and $\Omega_b = 0.0482754$, the Hubble parameter today is $H_0 = 67.556$ and the primordial power spectrum amplitude and tilt are $A_s= 2.215 \times 10^{-9}$ and $n_s = 0.9619$. {Note that we are currently setting the radiation component in the N-body simulation to $0$ in the background by imposing $\Omega_{\Lambda} = 1-\Omega_c-\Omega_b$ as well as at the level of the perturbations.} 
The fundamental mode of the comoving box is $k_f=8 \times 10^{-4}\, h/$Mpc. It corresponds to a box of $\sim 11.6\,$Gpc quite close to the cosmological horizon today $\sim 12.7\,$Gpc. The initial conditions are computed following Ref.~\cite{Montandon:2022ulz} with a custom branch of \texttt{MonofonIC}\footnote{The main branch is available on \href{https://bitbucket.org/ohahn/monofonic/src/master/}{https://bitbucket.org/ohahn/monofonic/src/master/} and the modified branch on \href{https://bitbucket.org/tomamtd/monofonic/src/relic/}{https://bitbucket.org/tomamtd/monofonic/src/relic/}.}. We use a grid of $1024$ points thus the Nyquist frequency limiting the resolution of \texttt{gevolution} is $0.41\,h/$Mpc. Each particle have a mass of $5.15 \times 10^{12}\, \mathrm{M}_{\odot}/h$. The bottleneck to the formation of dark matter halos in this numerical setup is the grid resolution that would need to be an order of magnitude larger. Having such a large grid is very computationally challenging and moreover workarounds already exist on the market in the form of tree algorithms \cite{Springel:2020plp} or of multi-grid approaches \cite{Teyssier:2001cp}.

The displacement field, and therefore the initial conditions, is computed with \texttt{gevolution} at redshift $50$ by using the second-order discrete Lagrangian method developed in Ref.~\cite{Adamek:2021rot}\footnote{The branch of gevolution containing the second-order initial conditions can be found on \href{https://github.com/TomaMTD/gevolution-1.2}{https://github.com/TomaMTD/gevolution-1.2}.}. The general relativistic initial conditions encompass all the terms up to second order in perturbation theory in Poisson gauge including radiation, while the Newtonian initial conditions are computed in the N-body gauge \cite{Fidler:2015npa}, see Ref.~\cite{Adamek:2021rot} for more details.

In order to populate all the modes of the Fourier grid in the \texttt{gevolution} simulations, we use $2048^3$ particles. The exact same Newtonian initial conditions are used for the Newtonian simulations performed with \texttt{gevolution} and \texttt{Gadget-4}. For the simulation performed with \texttt{Gadget-4}, we use a softening length of $250\,$kpc$/h$. As we are simulating relativistic effects along with halo formation, our \texttt{Gadget-4} simulation spans four orders of magnitudes in scales and is the computational bottleneck of this whole simulation pipeline. Indeed, one \texttt{gevolution} simulation takes $\sim 1000$ CPUh while the corresponding \texttt{Gadget-4} simulation costs $0.5$ MCPUh.

\section{Matter density statistics}\label{sec:stats}
In this section, we compare the matter density power spectrum and bispectrum of the different simulations: \texttt{Gadget-4}, \texttt{gevolution} in Newtonian mode labelled ``\texttt{gevolution} Newton'', \texttt{gevolution} in GR mode labelled ``\texttt{gevolution} GR'' and the combination of the different simulations according to \eqref{eq:combined}. 
The matter density contrast is defined as $\delta_m(\boldsymbol x)=\rho_m(\boldsymbol x)/\bar \rho - 1$ where $\rho_m$ is the matter density and $\bar \rho$ is the mean density of the Universe. {To evaluate the density, we use the counting density as computed by the code \texttt{Pylians3} \cite{Pylians}. For the Newtonian simulations \texttt{Gadget} and ``\texttt{gevolution} Newton'', this corresponds to the physical density. For ``\texttt{gevolution} GR'', the counting density cannot be directly interpreted as the physical density because of the volume perturbation of the Poisson gauge. However, the combined simulations are also in Poisson gauge such that they can be compared at large scales with ``\texttt{gevolution} GR''.}

\subsection{Matter power spectrum}
\label{sec:powerS}
We start by comparing the matter power spectrum and bispectrum of the different simulations. The power spectrum of two isotropic and homogeneous fields $\delta_i$ and $\delta_j$ reads   
\begin{equation}\label{eq:powerspectrum}
    \left< \delta_i (\boldsymbol{k}_1) \delta_j (\boldsymbol{k}_2)\right> = (2\pi)^3 \delta_{\mathrm D}(\boldsymbol{k}_1+\boldsymbol{k}_2) P_{ij}(k_1)\,,
\end{equation}
where $\delta_{\mathrm D}$ is the Dirac distribution and $\left< \dots \right>$ is the average over the modes.

\begin{figure}
    \centering
    \includegraphics[scale=0.41]{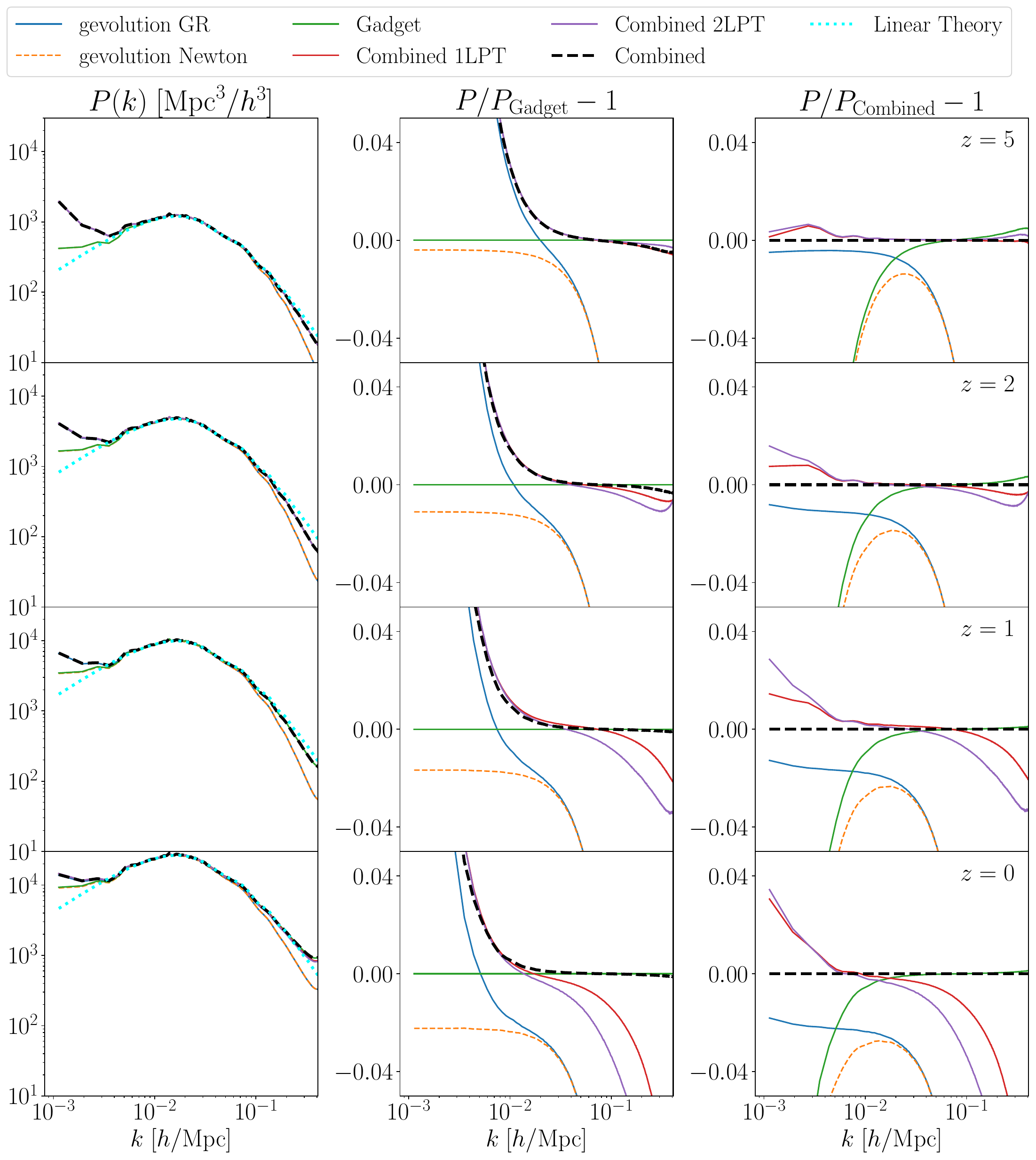}
    \caption{Power spectra for 4 redshifts. Blue: GR simulation (\texttt{gevolution} GR), orange: Newtonian one (\texttt{gevolution} Newton), green: \texttt{Gadget-4} one, black: combination of simulations with Eq.~\eqref{eq:combined}, red and violet: combination of \texttt{Gadget-4} with 1LPT and 2LPT, cyan: linear power spectrum in the synchronous gauge from \texttt{CLASS}. The second (third) column shows the relative difference with the reference \texttt{Gadget-4} (\texttt{gevolution} GR) simulation. The combined simulation is able to retrieve the relativistic effects at large scales as well as the nonlinear behaviour at small scales.}
    \label{fig:Pk}
\end{figure}

In the first column of Fig.~\ref{fig:Pk}, we show the matter density contrast power spectrum $P_{mm}$ for $4$ redshifts $z\in[5, 2, 1, 0]$. The second and third columns show the relative difference with the \texttt{Gadget-4} and the relativistic \texttt{gevolution} simulations respectively. The colours indicate the different simulations and combinations, see caption of Fig.~\ref{fig:Pk}.

\paragraph{Small-scale behaviour.} As expected, 
{the power spectra of the gevolution simulations (relativistic and Newtonian) are equal at small scales}. Indeed, deep inside the horizon, relativistic effects are negligible. However, we can see that the small-scales nonlinearities are underestimated in both simulations performed with \texttt{gevolution} compared to the \texttt{Gadget-4} simulations. This is due to the particle mesh nature of the code that becomes less accurate close from the Nyquist frequency. Conversely, the tree algorithm of \texttt{Gadget-4} allows to resolve Newtonian interactions down to the force softening scale. 

\paragraph{Large-scale behaviour.}
At large scales, we observe the standard behaviour due {the gauge dependent relation between gravitational potential and density}. The relativistic simulation has an increasing power due to the Poisson gauge. In the linear theory without radiation, this is a pure fictitious gauge effect that cannot be observed. The Newtonian simulations have the same expected trend, but \texttt{gevolution} still underestimates the large scales perturbation power, as one can see in dashed orange curve in the middle panel. This difference increases as the redshift decreases. In Ref.~\cite{Quintana-Miranda:2023eyn}, this behaviour is observed and attributed to the first-order gradient computation implemented in \texttt{gevolution}. {A second-order gradient computation is implemented and can reduce this large-scale offset. This would however even more degrade the small-scale power.} The same loss of power should affect in the same way the large scales of the relativistic simulation \texttt{gevolution} GR. 

\paragraph{Combined simulations.}
With the black dashed line, we show the power spectrum of the dark matter density perturbations computed after combining the three simulations according to Eq.~\eqref{eq:combined}. The small-scale Newtonian perturbations are recovered with a relative difference of $\sim 0.1\%$ for $z<2$ and $\sim 0.5\%$ for $z\geq2$. At large scales, the relative difference to the \texttt{gevolution} simulation is less than $\sim 2$\%, reaching $0.5\%$ for $z=5$. The numerical inaccuracy due to the first-order gradient computation discussed in the previous paragraph partially cancels in the subtraction of Eq.~\eqref{eq:xiGR}. Indeed, we can see that ``\texttt{gevolution} GR'' in the third column underestimates in a very similar manner the power at large scales compared to the ``combined'' simulation. The power is however underestimated with respect to LPT and this discrepancy amplifies as the redshift decreases. That shows that the cancellation is not perfect. {This non-perfect cancellation might be due to the fact that the error introduced by neglecting linear radiation is gauge dependent, see Fig.~1 of Ref.~\cite{Adamek:2017grt}.}

\paragraph{Comparison to the combination with LPT} The LPT results are of the order of $1\%$ to $10\%$ compared to the \texttt{Gadget-4} simulation at small scales. Since the gauge transformation should not affect the small scales, this indicates theoretical or numerical errors in the evaluation of the fields. Surprisingly, while for redshift $z=5$, 2LPT becomes better than 1LPT, for $z<5$, 1LPT shows a better agreement than 2LPT. This may indicate that perturbation theory has broken down. Indeed, the method employed and described in Ref.~\cite{Adamek:2021rot} is perturbative and may therefore not be accurate enough at low redshift/small scales. Moreover, on the theoretical side, we neglect linear radiation in the dynamics of our simulations while it is included in the LPT computation. This theoretical error changes the size of the cosmological horizon, and hence the scale of the large-scale turn over in the Poisson gauge. Therefore, as already observed in Ref.~\cite{Hahn:2016roq}, this would cause a large error at large scales because the power spectrum diverges in this gauge for modes smaller than the horizon. This would explain the large-scale discrepancy between the ``combined'' simulation and LPT visible in the third column.

\subsection{Matter bispectrum}
\begin{figure}
    \centering
     \includegraphics[width=\textwidth]{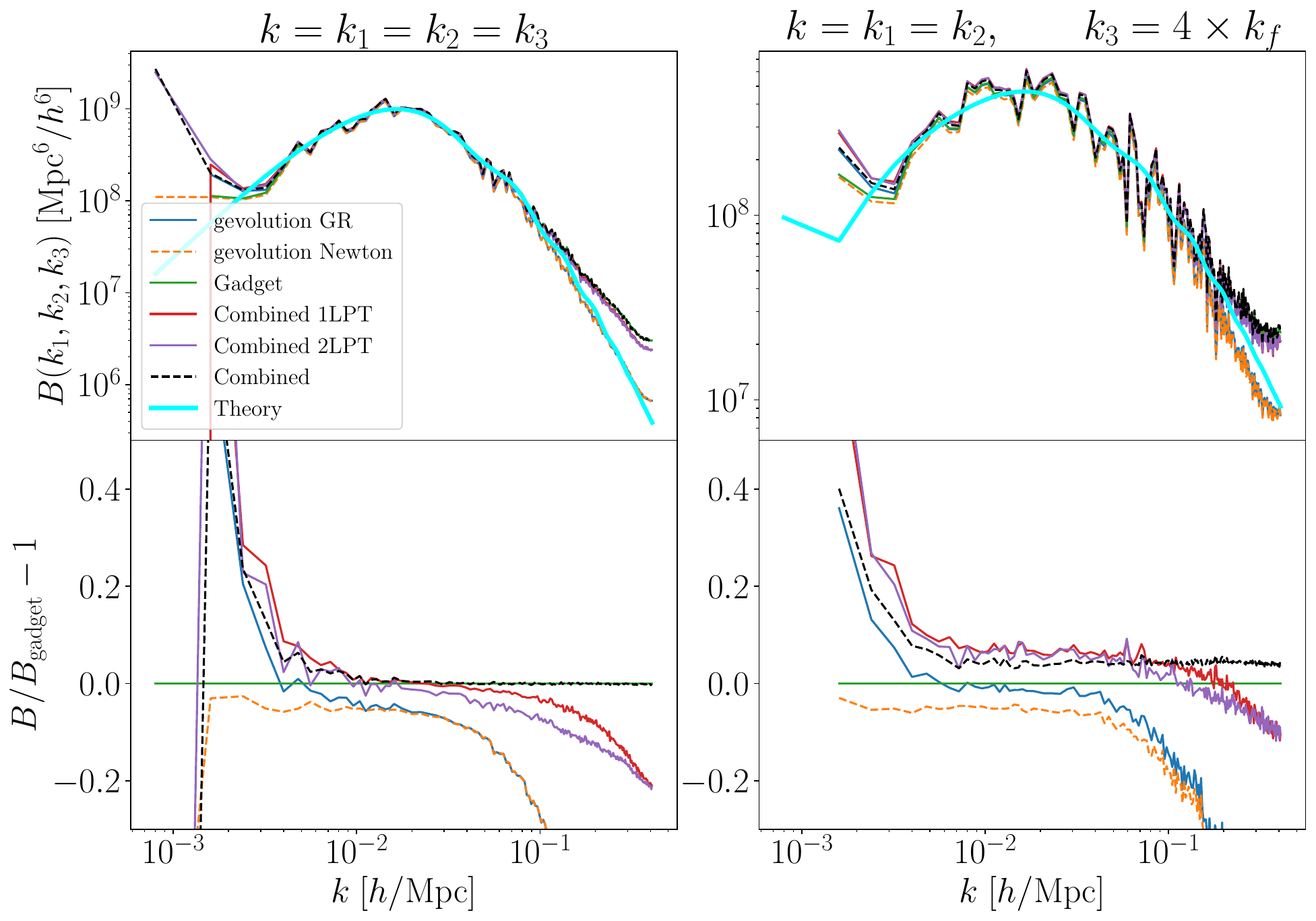}
    \caption{Matter bispectrum at redshift $0.1$ for two different triangle configurations: equilateral in the left panel and squeezed (one modes fixed to four times the fundamental mode of the box, i.e. $3.2\times 10^{-3}\, h/$Mpc) in the right panel. 
    The colours are the same as used in Fig.~\ref{fig:Pk}. The theory, in cyan, shows the Newtonian tree-level bispectrum, see Refs.~\cite{Tram:2016cpy, Villa:2015ppa}.
    The second row shows the relative difference with the reference \texttt{Gadget-4} simulation.}
    \label{fig:Bk}
\end{figure}

The bispectrum of the matter density field $\delta_m$ is defined as  
\begin{equation}\label{eq:bispectre}
    \left< \delta_m (\boldsymbol{k}_1) \delta_m (\boldsymbol{k}_2) \delta_m (\boldsymbol{k}_3)\right> = (2\pi)^3 \delta_{\mathrm D}(\boldsymbol{k}_1+\boldsymbol{k}_2 + \boldsymbol{k}_3) B_{mmm}(k_1, k_2, k_3)\,.
\end{equation}
In Fig.~\ref{fig:Bk}, we show in the first row the bispectrum for two different triangle configurations at redshift $0.1$. In the left panel, the equilateral configuration and on the right squeezed triangle with the two modes $k_1 = k_2$ varying and the third mode $k_3$ fixed to four times the fundamental frequency of the box. The colours used are the same as in Fig.~\ref{fig:Pk}. To significantly improve the bispectrum measurements which would be otherwise highly noisy, we have used the pairing method \cite{Angulo:2016hjd}.

Many of the results that we have found for the power spectrum in subsection \ref{sec:powerS} also hold for the bispectrum. At small scales, both \texttt{gevolution} simulations underestimate the bispectrum because of the particle mesh nature of the code.
At large scales, the Newtonian simulation performed with \texttt{gevolution} underestimates the bispectrum because of the first-order gradient computation. The combination of the three different simulations, called ``combined'' in the legend, has the advantage of the \texttt{Gadget-4} simulation at small scales, with the same $0.1\%$ relative difference as in the power spectrum, with the trend of the relativistic simulations at large scales with slightly more amplitudes which comes from the cancellation of the gradient numerical errors in the GR displacement field computation in Eq.~\eqref{eq:xiGR}. This is particularly visible when looking at the squeezed limit configuration. The combination of \texttt{Gadget-4} with LPT shows a loss of power at the level of  $10$ to $20\%$ with respect to \texttt{Gadget-4}, similar to the one observed in the power spectrum. At large scales, the amplitude of the relativistic trend is again larger than the one of the ``combined'' simulation, which shows that the cancellation of the numerical error is partial. 
Since one mode is fixed to a large scale $k_3=4\times k_{\rm f}$, we observe a constant offset (except for the largest scales) between the Newtonian and the relativistic simulations performed by \texttt{gevolution}. When we combined the simulations' positions, we recover a similar constant offset, but now compared to \texttt{Gadget-4}. As expected, this offset is even larger for LPT at large scales, which then loses power at small scales.

In this section, we have shown that the combination of the particle positions defined in Eq.~\eqref{eq:combined} produces dark matter power spectrum and bispectrum that have the advantages of the \texttt{Gadget-4} and of the relativistic \texttt{gevolution} simulations. At small scales, the perturbations are not limited by the Nyquist frequency and the power spectrum and bispectrum of the combined simulations have a relative difference with the \texttt{Gadget-4} simulation of $0.1\%$ for $z<2$. At large scales, the combined simulation has the same relativistic behaviour as the relativistic simulation, with an offset similar to the one observed also at large scales between the Newtonian simulations. The comparison with the large scales results obtained with LPT show that the errors introduced by the first-order gradient computation of \texttt{gevolution} partially cancel. 

The combination of simulations shows overall a good agreement with LPT at high redshift ($z=5$). However, the combination of \texttt{Gadget-4} and LPT degrades the power at small scales for the lower redshifts. This might be due to the iterative method used in this article to generate the displacement fields which is a perturbative numerical method and therefore might break down at low redshift and small scales. Finally, note that, as demonstrated by Ref.~\cite{Quintana-Miranda:2023eyn}, improving the density field accuracy also leads to a better precision of all metric potentials $\phi$, $\chi$, $B_i$ and $h_{ij}$. 
We now move to the study of the statistics of the halo formed in our simulations. Given the fact that the combination of \texttt{Gadget-4} with LPT is inaccurate at small scales, we will now only focus on the comparison of the combination of the simulations with \texttt{Gadget-4}.

\section{Halo mass function and halo bias} \label{sec:halos}
Dark matter halos are a biased tracer of the underlying dark matter distribution, see Ref.~\cite{Desjacques:2016bnm} for a review on the galaxy bias. The scale dependence of the bias in general relativity was studied in Refs.~\cite{Challinor:2011bk, Bruni:2011ta, Baldauf:2011bh, Jeong:2011as, Bertacca:2015mca, Umeh:2019qyd,Calles:2019prs}. The derivation of the galaxy density up to second order in any gauges was in particular performed in Ref.~\cite{Umeh:2019qyd}. Recent numerical studies have been conducted on halo bias in a relativistic context \cite{Lepori:2021lck,Lepori:2022hke}. The authors used \texttt{gevolution} to simulate $5760^3$ N-body particles in a volume of $(4032$ Mpc$/h)^3$. Their resolution in mass was $3\times 10^{10} \;M_\odot/h$ but their halo formation is limited by the particle mesh to $700\;$kpc$/h$. The method presented in our article can be used to bypass this bottleneck to form halos. We are indeed numerically following a box almost two times larger than in Ref.~\cite{Lepori:2021lck} with a spatial resolution of $250\;$kpc$/h$, our simulations are only limited by the number of particles forming halos. Moreover, the box size used in Ref.~\cite{Lepori:2022hke} is small enough to neglect effects due to the horizon. In this section, we study the halo field and in particular the impact of horizon scales on them. Note that our result are only valid in the Poisson gauge. Heading toward a gauge invariant prediction requires to combine our results with a ray tracing algorithm as in Ref.~\cite{Montandon:2022ulz}.

\subsection{Halo Mass Function}

\begin{figure}
    \centering
    \includegraphics[width=\textwidth]{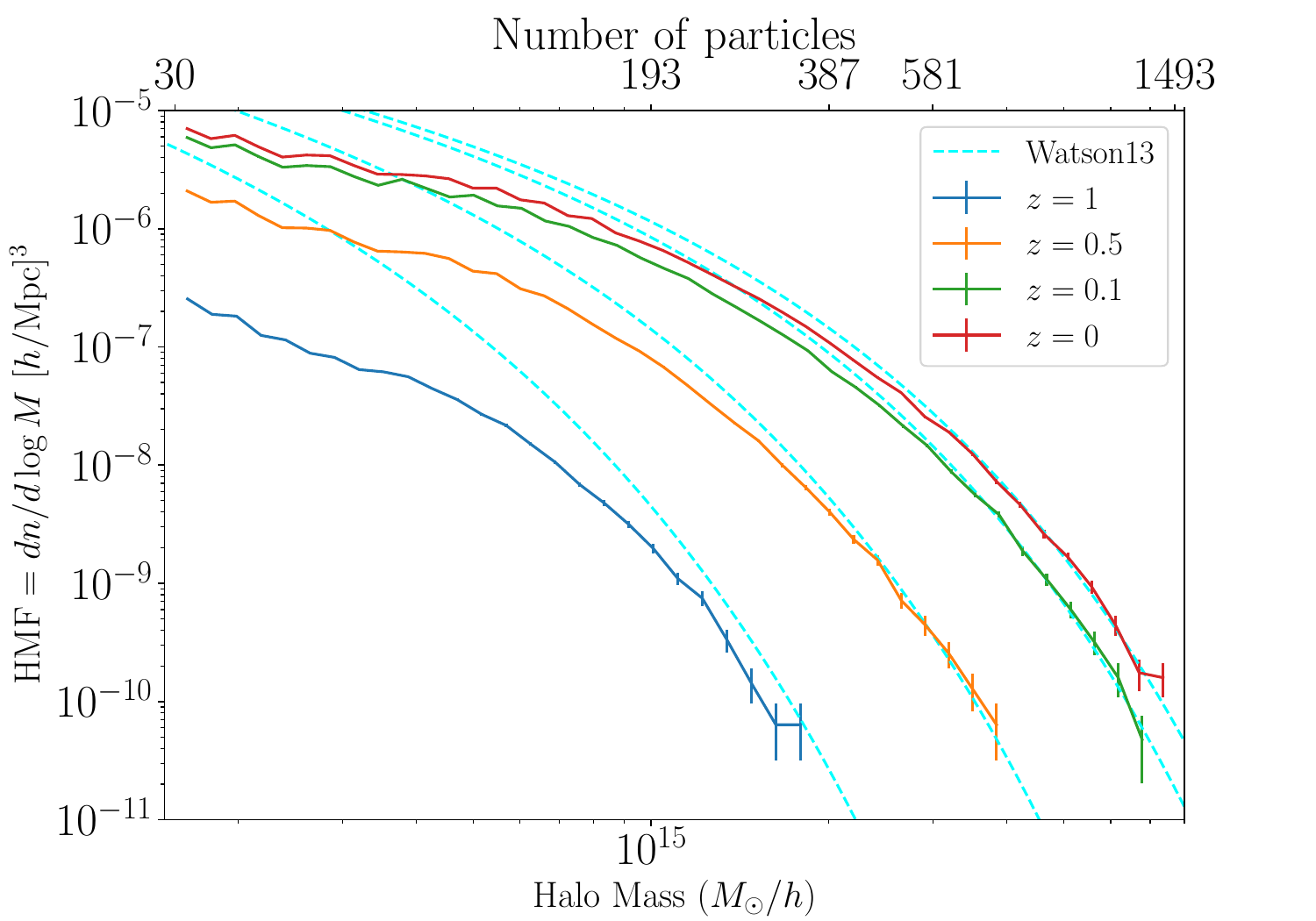}
    \caption{Halo mass function of the \texttt{Gadget-4} simulation for $4$ different redshifts. Dashed lines indicate the models of Refs~\cite{Watson:2012mt} computed with \texttt{Colossus}. For low masses, the halos are poorly resolved and the HMF is underestimated compared  to the theoretical prediction.} 
    \label{fig:hmf}
\end{figure}
In Fig.~\ref{fig:hmf}, we display the Halo Mass Function (HMF) for 4 redshifts. From the bottom to the top, we have $z=1$ in blue, $z=0.5$ in orange, $z=0.1$ in green, and $z=0$ in red. The range of mass between $2 \times 10^{14} \;M_\odot/h$ to $9.0 \times 10^{15} \;M_\odot/h$ gives halos with a least $30$ particles. In dashed lines, we add the model of Ref.~\cite{Watson:2012mt} plotted with \texttt{Colossus} \cite{Diemer:2017bwl}. We stress that all HMF models calibrated with N-body simulations have to be extrapolated to meet the very massive halos that we are simulating. These halos are well converged for masses larger than $\sim 2\times 10^{15} \;M_\odot/h$. Hence, the HMF is underestimated for all masses at $z=1$. The information on the HMF is now useful to measure the halo bias. Those measures need a trade-off between statistics (number of halos studied) and the convergence of our simulations. As we will see in the next section, keeping all halos with more than $30$ particles gives a reasonable measure of the halo bias. 

\subsection{Linear halo bias}
In Newtonian gravity, and in the absence of primordial non-Gaussianity, the linear halo density contrast is proportional to the matter density contrast: $\delta_h^{\mathrm{N}} = b^{\mathrm{N}}_1 \delta^{\mathrm{N}}_m$. In GR, a gauge effect leads to an additional relativistic term proportional to $\mathcal H^2 / k^2$: 
\begin{equation}\label{eq:b1}
    \delta^{(1)}_h(z, \boldsymbol{k}) = \left(b_1(z)  + b^{\mathrm {GR}}_{1}(z) \frac{\mathcal H^2}{k^2} \right) \delta^{(1)}_m(z, \boldsymbol{k}) \,.
\end{equation} 
{Recall that we measure the counting density. In the Newtonian simulation, it matches the physical density $\hat \delta$. In Poisson gauge, its relation with the physical density reads at linear order
\begin{equation}\label{eq:physical_density}
    \delta^{(1)}_{X, {\rm P}} = \hat \delta^{(1)}_{X, {\rm P}} - 3 \phi^{(1)} \,,
\end{equation}
where $X$ stands either for $m$ (matter) or $h$ (halos) and $\phi$ is the gravitational potential. 
}
Using Eq.~(53) of Ref.~\cite{Umeh:2019qyd} at linear order along with Eqs.~(5.18) and (6.8) of Ref.~\cite{Villa:2015ppa}, the linear halo bias in the Poisson gauge for the counting density reads:
\begin{equation}\label{eq:bgr}
    b^{\mathrm {GR}}_{1}(z) = 3 \left(f(z) + \frac{3}{2} \Omega_m(z) \right) (1-b_1(z)) \,,
\end{equation}
where $f$ is the linear growth rate. In the Newtonian Gaussian case, we theoretically expect $b^{\rm GR}_1= 0$, but PNG induce a scale-dependent bias \cite{Dalal:2007cu} that is now pretty standard to constrain in the analysis of the current large observational surveys. {Note also that the constant bias in Newtonian simulation can be interpreted in a GR context \cite{Fidler:2018geb}.}

\begin{figure}
    \centering
    \includegraphics[scale=0.33]{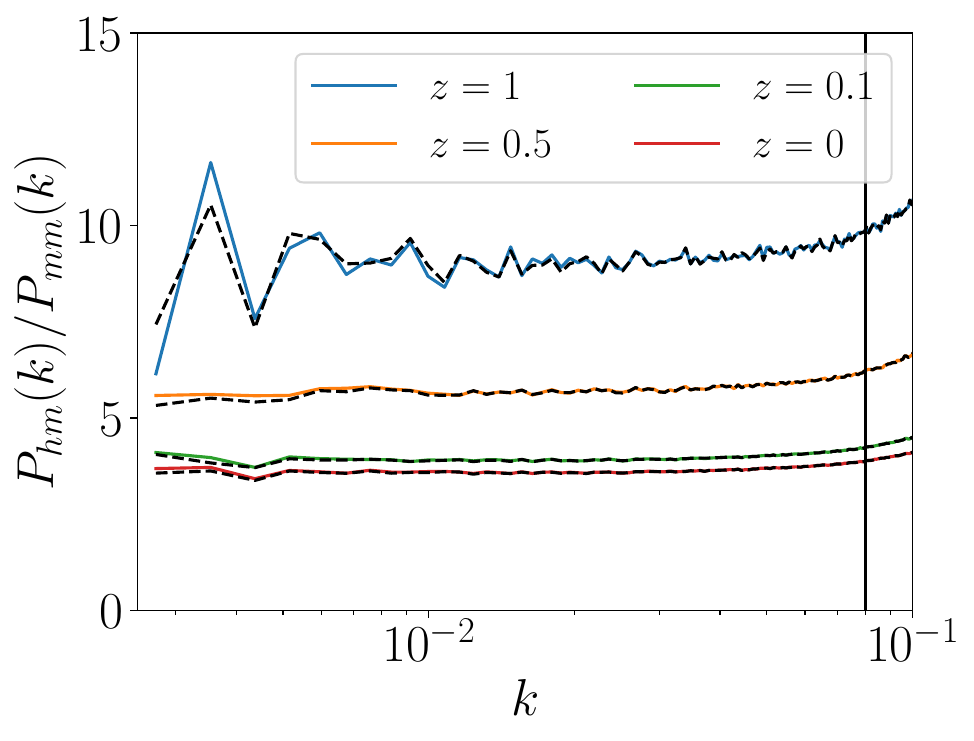}
    \includegraphics[scale=0.33]{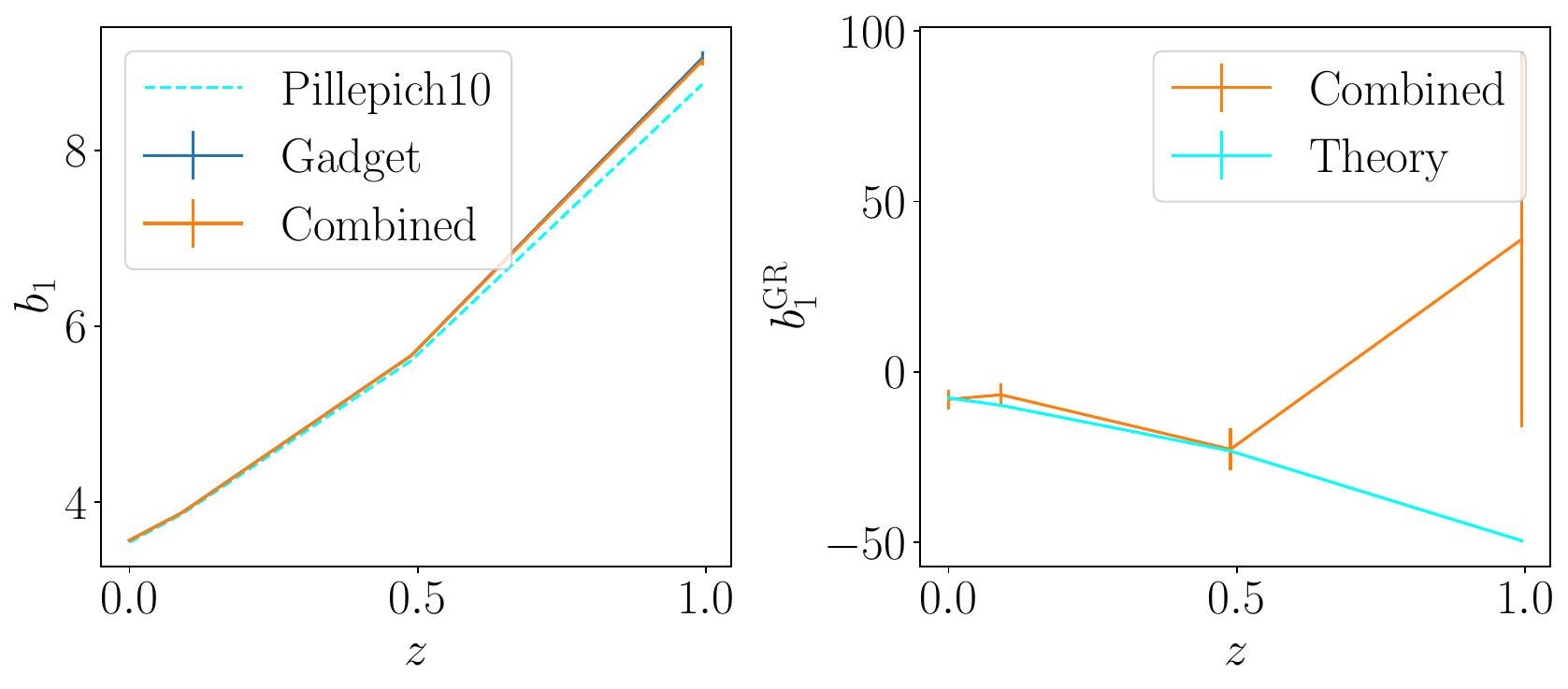}
    \caption{Left panel: Ratios of the halo-matter cross-spectrum and the matter power spectrum for four redshifts: $1$, $0.5$, $0.1$ and $0$. $b_1$ and $b_1^{\mathrm{GR}}$ are fitted with Eq.~\eqref{eq:b1} for $k<k_{\rm cut}=0.08\, h$/Mpc. The cutoff $k_{\rm cut}$ is indicated as a black vertical line. Middle  and right panels: fitted bias parameters $b_1$ and $b_1^{\rm GR}$ as a function of the redshift. The cyan curves show the theoretical predictions of Ref.~\cite{Pillepich:2008ka,Umeh:2019qyd}. For $z=0$, $z=0.1$, we have detected for the first time a negative linear relativistic halo bias. 
    }
    \label{fig:bias}
\end{figure}

In the left panel of Fig.~\ref{fig:bias}, we show $b_1(z, k)$ as a function of $k$. The colours (from top to bottom) blue, orange, green and red indicate the four different redshifts $1$, $0.5$, $0.1$ and $0$. As expected, as we go to large scales, the ratio $b_1(z, k)$ tends to a constant. This constant itself scales like the redshift. 
The black dashed lines almost superposed to the coloured curves show the bias computed with the ``combined'' simulation. They follow exactly the \texttt{Gadget-4} simulation at small scales, and seem to have a decreasing amplitude with respect to the \texttt{Gadget-4} simulation as we go to large scales. To estimate the deviation, we fit the largest scales up to the cut-off indicated in black vertical lines $(k_{\rm cut} = 0.08\;h/$Mpc$)$. The small scales nonlinearities induce an increase of $P_{hm}/P_{mm}$ that we take into account by adding a term $\propto k^2$ to our fitting procedure with Eq.~\eqref{eq:b1}. This allows us to fit $b_1$ in a larger range of $k$ \cite{Barreira:2021ueb}. A more involved treatment would feature non-linear bias parameters \cite{Desjacques:2016bnm}.

The fitted linear bias $b_1$ can be read in Table \ref{tab:b1} and is shown as function of the redshift in the middle panel of Figure \ref{fig:bias}. {Since the halo mass function is underestimated, see Fig.~\ref{fig:hmf}, the bias will also be underestimated. To correct this effect, we use the method of ``abundance matching'' \cite{Euclid:2022qde}.} 
The mismatch between HMF fits and our measurements in Fig.~\ref{fig:hmf} due to numerical resolution effects can be either interpreted as an underestimation of the abundance at fixed mass, or an underestimation of the mass at fixed abundance. Interpreting it as the latter allows us to recover the expected result by abundance matching, i.e. by reassigning halo masses to recover the expected abundance given by the fit curve of Ref.~\cite{Watson:2012mt}. 
The results obtained in the \texttt{Gadget-4} and the ``combined'' simulations are respectively shown in solid blue and orange lines. We do not observe significant impact of relativistic effects on the linear Newtonian bias estimation.  
The dashed lines show the bias model of Ref.~\cite{Pillepich:2008ka} obtained with \texttt{Colossus} and integrated over the mass range of interest:
\begin{equation}
    b_1^{\rm model}(z) = \frac{\int d M b_1(M, z)\frac{dn}{d\log M} }{\int d M \frac{dn}{d\log M}}\,.
\end{equation}
Note that the bias model that we have chosen \cite{Pillepich:2008ka} has been calibrated on simulations and has a range of mass validity up to $5 \times 10^{15}\;M_\odot/h$. Our mass functions goes slightly beyond this limit but our estimated Newtonian linear bias is well compatible. 

\begin{table}[]
    \centering
    \begin{tabular}{|c||c|c||c|c|}
\hline
         Simulations & \texttt{Gadget-4} & \multicolumn{2}{|c|}{\texttt{Combined}} \\
\hline
     Bias     & $b_1$  & $b_1$ & $b_1^{\rm GR}$\\
\hline
\hline
$z = 1$ & $9.047 \pm 0.077 $ &
$9.017 \pm 0.057 $ & $38.848 \pm 55.141 $ \\
\hline
$z = 0.5$ & $5.667 \pm 0.007 $ &
$5.667 \pm 0.007 $ & $-22.658 \pm 6.165 $ \\
\hline
$z = 0.1$ & $3.880 \pm 0.005 $ &
$3.879 \pm 0.005 $ & $-6.693 \pm 3.453 $ \\
\hline
$z = 0$ & $3.566 \pm 0.004 $ &
$3.565 \pm 0.004 $ & $-8.054 \pm 2.840 $ \\
\hline
    \end{tabular}
    \caption{Halo bias obtained by fitting Eq.~\eqref{eq:b1} to the large scales ($k<k_{\rm cut}=0.08\, h/$Mpc) of the first panel Fig.~\ref{fig:bias}. We detect a non-zero GR halo bias at more than $3\sigma$ for all redshifts.}
    \label{tab:b1}
\end{table}

While, we expect a vanishing $b_1^{\rm GR}$ in the \texttt{Gadget-4} simulation, the residual cosmic variance, noise and shot noise that underestimates the halo mass function, produce a significant large systematic effect. To estimate these effects, we measure $b_1^{\rm GR}$ in the Newtonian simulations and subtract it from the measure of $b_1^{\rm GR}$ in the ``combined'' simulation. {Moreover, to test the robustness of our measurements, we compare the measurements obtained by removing the large scale points. We find that the measurements become stable and consistent with the theory after removing the first two points.} 
The result obtained is shown in the right panel of Fig.~\ref{fig:bias} in orange line. The cyan line show the theoretical result of Eq.~\eqref{eq:bgr}. Note that we use the estimated linear bias, which means that the theoretical curve has small error bars almost invisible on the plot. The three lowest redshifts, which have much more statistics and have a larger converged range of mass for the halo mass function are well centred on the theoretical prediction and exclude both $0$ at more than $2\sigma$ to $3.6\sigma$, see Table~\ref{tab:b1} for the exact values. For redshift $z>0.5$, the relativistic halo bias $b_1^{\rm GR}$ has larger error bars and include $0$ at $\sim 1\sigma$ and the theoretical prediction at $\sim 1.5\sigma$.   

We have here measured for the first time the linear halo bias in a general relativistic context and isolated the GR bias contribution that is degenerate with PNG. In the next subsection, we propose some elements for the analysis of the halo bispectrum.

\subsection{Nonlinear halo bias}

\begin{figure}
    \centering
    \includegraphics[scale=0.5]{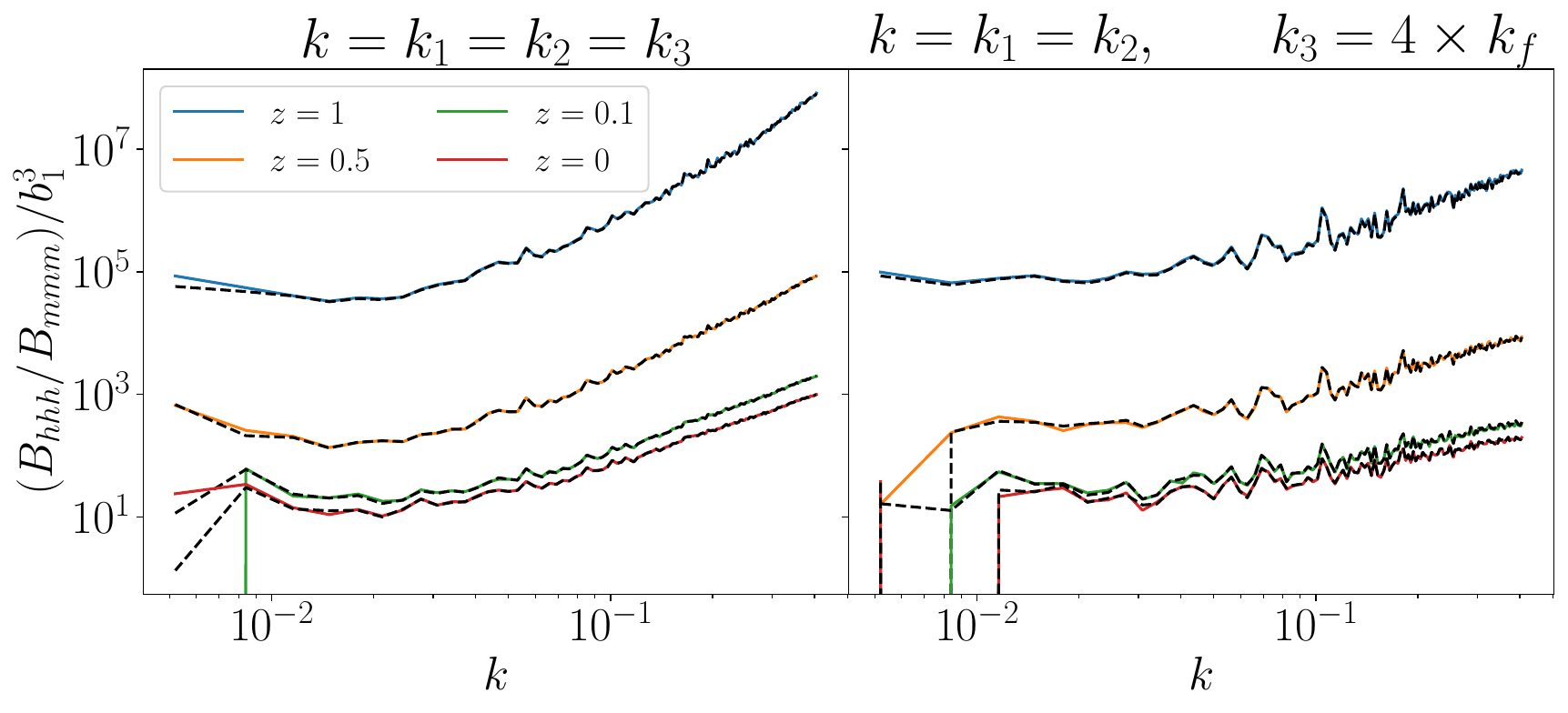}
    \caption{In the left (right) panel, $B_{hhh} / B_{mmm}$ in equilateral (squeezed) configurations with $4 \times k_f= 3.2\times 10^{-3}\, h/$Mpc. 
    The measurements obtained with \texttt{Gadget-4} are shown in colours while the black dashed lines are obtained with the combined simulations. The overall amplitude increases with the redshift. 
    }
    \label{fig:bias2}
\end{figure}

At second order, the equation for the halo density in Poisson gauge is more involved and requires two additional bias parameters: $b_2$ and $b_s$ eg.~\cite{Umeh:2019qyd}. Hence, the function $B_{hhh} / B_{mmm}$ is in principle a non-trivial function of the three bias parameters $b_1$, $b_2$ and $b_s$, and of the three different modes $k_1$, $k_2$ and $k_3$. A full analysis of the second-order bias would require a fit of the bispectrum including all the general relativistic terms, as well as all the second-order bias parameters, see for example Ref.~\cite{Lepori:2022hke}. As a first step, we show in Fig.~\ref{fig:bias2} the measurements obtained by dividing the halo bispectrum by the matter bispectrum. Similarly to Fig.~\ref{fig:bias}, the solid coloured lines represent the measurements performed in the \texttt{Gadget-4} simulation, and the dashed black lines almost superposed with the coloured lines are the corresponding measurements performed in the ``combined'' simulation. In the left panel, we present the equilateral configuration while in the right panel we have fixed one of the mode to $k_3=4\times k_{\rm f}$ and varied the two other modes such that large $k$'s correspond to squeezed configurations. With the accuracy of our simulation, we do not observe any significant difference between the relativistic and the Newtonian simulation. Thus, within our framework, we conclude that most of the relativistic effects impacting a measurement of $f_{\rm NL}$ would be present at the level of the scale dependant bias and not the halo bispectrum.

\section{Conclusion}\label{sec:conclusion}
In this paper, we have run and analysed the first simulation able to resolve both scales close to the cosmological horizon and halos in a general relativistic framework. To do so, we combined the desirable features of two N-body codes on the market: \texttt{gevolution} to compute the general relativistic displacement field and \texttt{Gadget-4} to resolve the halos thanks to its tree algorithm. We have numerically followed a box of $11.6$~Gpc with softening length $\epsilon=250$~kpc$/h$ thus simulating structure formation over four orders of magnitudes in length.  

Comparing the matter power spectrum and bispectrum between the different simulations, we demonstrated that the ``combined'' simulation obtained converges at small scales to the \texttt{Gadget-4} simulation, with a relative difference of $\sim 0.5\%$ for all redshifts. The combination of \texttt{Gadget-4} with a general relativistic displacement field evaluated with $1$- and $2$LPT shows a large loss of power at small scales due to the perturbative method employed to evaluate the displacement fields. 
At large scales, the ``combined'' simulation contains the same relativistic trend as \texttt{gevolution}, with an additional offset which comes from a cancellation of the errors generated by the first-order gradient computation in \texttt{gevolution}. The combination of \texttt{Gadget-4} with LPT shows a slightly larger offset, demonstrating that the cancellation is partial. 

Finally, we performed an analysis of the halos generated in our simulations by comparing the ``combined'' simulation with its \texttt{Gadget-4} counterpart. We studied the halo mass function for 4 redshifts between $z=1$ and $z=0$ to show that our simulations are only well converged for halo masses $\gtrsim 2\times 10^{15} \;M_\odot/h$. Using all halos with more than $30$ particles and the ``abundance matching'' method,  we have measured for the first time in a relativistic simulation the linear halo bias and found agreement with the theoretical prediction in the Poisson gauge \cite{Umeh:2019qyd}. We estimated the general relativistic scale dependence of the bias for the three lowest redshifts studied $z=0$, $z=0.1$ and $z=0.5$. Finally, we show the measurements of the halo bispectrum divided by the matter bispectrum, which naively corresponds to a second-order bias estimation. While at the level of the power spectrum, we were able to isolate the relativistic pollution to a measure of scale-dependent bias in $b_1$ (see Table \ref{tab:b1}), we did not observe any significant difference between the ``combined'' and the \texttt{Gadget-4} simulation at the level of the bispectrum.

    The prescription discussed in this work offers a simple way to include GR effects in Newtonian simulations. Our method is nonlinear, not restricted to the scalar sector and can also be extended to other kind of effects such as radiation, neutrinos or modified gravity. Forming halos in such general relativistic simulation complement the developments of Refs.~\cite{Fidler:2015npa,Umeh:2019qyd,Barrera-Hinojosa:2019mzo,Castorina:2021xzs,Elkhashab:2021lsk,Montandon:2022ulz,Quintana-Miranda:2023eyn,Pardede:2023ddq,Fidler:2018geb}, aiming on the long term to run a large relativistic simulations following eg.~\cite{Guandalin:2020snp,Coulton:2022qbc,Anbajagane:2023wif,Adame:2023nsx,Hadzhiyska:2024kmt}. Such large relativistic simulations will be out of paramount importance to validate the analysis pipeline for the observational data and to explore theoretical models and observational aspects in a controlled environment. For these analysis however, it will be crucial to compute actual observables such that the final results in the relativistic simulation becomes gauge invariant, as we have done for dark matter in Ref.~\cite{Montandon:2022ulz}. 
    In particular, this method needs to be adapted to a light-cone analysis. This can in principle be performed by using the same operation but with the four-dimensional positions of particles. The new positions of the particles can then be used for the ray tracing. This would allow us to reach high accuracy such as in Ref.~\cite{Breton:2021htu}, but in a gauge consistent and full relativistic framework such as in Ref.~\cite{Lepori:2022hke}. {Our method could also be combined with the code \texttt{LIGER} \cite{Borzyszkowski:2017ayl} which is able to post-process Newtonian simulations to produce the distribution of tracers on the light cone including linear relativistic effects.}
    The accuracy of such methods will be studied in future works.
    Toward obtaining gauge invariant observables, an intriguing question is whether a halo finder designed for Newtonian purpose needs to be adapted to the gauge adopted. These considerations will be essential not to discard the largest scales currently observed and to gain unparalleled insight concerning our Universe.

\acknowledgments
The simulation results presented have been achieved using the Vienna Scientific Cluster (VSC) and the Grand Equipement National de Calcul Intensif (GENCI). TM and OH especially thank Siegfried Hoefinger, Naima Alaoui, Alice Faure and Stephane Nou for many technical supports, and Vivian Poulin, Cornelius Rampf and Florian List for very helpful discussions. All authors thank Julian Adamek for reading the manuscript and for giving very useful comments. 
TM is supported by funding from the European Research Council (ERC) under the European Union’s HORIZON-ERC-2022 (grant agreement no. 101076865). CS acknowledges funding from the European Research Council (ERC) under the European Union's Horizon 2020 research and innovation program (grant agreement No.\ 834148).

\appendix

\bibliographystyle{JHEP.bst}
\bibliography{main.bib}

\end{document}